\documentclass[fleqn,usenatbib]{mnras}

\usepackage{newtxtext,newtxmath}
\usepackage[T1]{fontenc}
\usepackage{ae,aecompl}
\usepackage{hyperref}

\usepackage{graphicx}	% Including figure files
\usepackage{amsmath}	% Advanced maths commands
\usepackage{comment}
\usepackage{tabularx}

\newcommand{\kepler}{\textit{Kepler}}
\newcommand{\LUNA}{\texttt{LUNA}}

\newcommand{\cofiam}{\texttt{CoFiAM}}
\newcommand{\keras}{\texttt{Keras}}
\newcommand{\tensorflow}{\texttt{TensorFlow}}

\title[CNN Identification of Exomoons]{Identifying Potential Exomoon Signals with Convolutional Neural Networks}

\author[Teachey \& Kipping]{
Alex Teachey$^{1,2}$\thanks{E-mail: amteachey@asiaa.sinica.edu.tw} \&
David Kipping$^{2,3}$
\\
$^{1}$Academia Sinica Institute of Astronomy \& Astrophysics, Taipei, Taiwan R.O.C.\\
$^{2}$Department of Astronomy, Columbia University in the City of New York, USA\\
$^{3}$Center for Computational Astrophysics, Flatiron Institute, 162 Fifth Avenue, New York, NY 10010 USA
}

\date{Accepted 15 September 2021. Received 12 August 2021; in original form 23 October 2020}

\pubyear{2020}

\begin{document}
\label{firstpage}
\pagerange{\pageref{firstpage}--\pageref{lastpage}}
\maketitle

\begin{abstract}
Targeted observations of possible exomoon host systems will remain difficult to obtain and time-consuming to analyze in the foreseeable future. As such, time-domain surveys such as \kepler, K2 and TESS will continue to play a critical role as the first step in identifying candidate exomoon systems, which may then be followed-up with premier ground- or space-based telescopes. In this work, we train an ensemble of convolutional neural networks (CNNs) to identify candidate exomoon signals in single-transit events observed by \kepler. Our training set consists of ${\sim}$27,000 examples of synthetic, planet-only and planet+moon single transits, injected into \kepler\ light curves. We achieve up to 88\% classification accuracy with individual CNN architectures and 97\% precision in identifying the moons in the validation set when the CNN ensemble is in total agreement. We then apply the CNN ensemble to light curves from 1880 Kepler Objects of Interest with periods $>10$ days ($\sim$57,000 individual transits), and further test the accuracy of the CNN classifier by injecting planet transits into each light curve, thus quantifying the extent to which residual stellar activity may result in false positive classifications. We find a small fraction of these transits contain moon-like signals, though we caution against strong inferences of the exomoon occurrence rate from this result. We conclude by discussing some ongoing challenges to utilizing neural networks for the exomoon search.
\end{abstract}


%%%%%%%%%%%%%%%%%%%%%%%%%%%%%%%%%%%%%%%%%%%%%%%%%%

%%%%%%%%%%%%%%%%% BODY OF PAPER %%%%%%%%%%%%%%%%%%

\section{Introduction}
The identification of candidate exomoon signals in time-domain photometry remains a significant challenge. With such an enormous data volume to comb, and rigorous Bayesian model selection so computationally expensive \citep[e.g.][]{HEKV}, how does one efficiently identify potentially interesting signals? Increasingly, the astronomical community has turned to the use of machine learning for approaching such problems \citep[see a comprehensive review from][]{ML_review}. Indeed, mentions of the term in refereed publication abstracts has grown exponentially in the last decade, from about 200 articles in 2010 to more than 4000 in 2020 -- a growth rate far exceeding that of the yearly literature increase. In some applications, machine learning may be able to identify patterns that human beings cannot (in, for example, higher dimensional spaces where a system state is determined by more than three variables). In other contexts, machine learning algorithms may be only as good as what the human eye can discern (for example, computer vision), but are nevertheless employed for their speed (e.g. autonomous vehicles) and / or capacity for handling vast amounts of data (e.g. mail sorting).

Recent work \citep[e.g][]{shallue:2018, pearson:2018, dattilo:2019, yu:2019, osborn:2020} with the \kepler, K2, and TESS dataset have shown that convolutional neural networks (CNNs) can be effective in distinguishing genuine exoplanet transits from false positives. These efforts employed two light curve inputs -- a ``local'' and a ``global'' view --
and in some cases also incorporated additional observables as inputs, achieving up to 96\% accuracy and confirming a number of new planet candidates. 

\label{sec:introduction}

In most of the works highlighted above (the exception being \citealt{osborn:2020}), the networks were trained on a sample of previously confirmed systems. This is certainly a reasonable approach, since it incorporates the great diversity of systems so far discovered and avoids the requirement of producing simulated systems. At the same time, there are some downsides to this methodology, namely, that it 1) runs the risk of inadvertently teaching the networks about (and therefore encoding) the inherent biases of previous detection / confirmation methods, 2) limits the training sample to those systems that have already been detected, and 3) ground-truth values are not known with perfect precision or accuracy, adding noise to the training set. We note, however, that there is some evidence that CNNs can be robust against even high levels of label noise \citep{rolnick:2017}, making this last point not especially worrisome.

In this work, we apply an ensemble of CNNs to the problem of identifying candidate exomoon signals in single transit events. We ask the CNNs to make a simple binary classification: for a given light curve segment, is there a moon-like transit present, or not? Because we do not yet have an abundance of exomoon detections in the literature, we must train the CNNs using artificial light curves, being careful to make sure these light curves are as authentic as possible and represent the full range of possible system architectures and geometries. With our trained ensemble of CNN classifiers in hand, we go on to analyze 1880 Kepler Objects of Interest (KOIs) -- confirmed and candidate exoplanets -- in search of exomoons. We note that our implementation is appreciably different from that described by \citealt{alshehhi:2020}, who concurrently and independently also developed a CNN approach to the problem of identifying exomoons in time-domain photometry. In contrast to that work, we also go beyond proof of concept to apply our classification network to the \textit{Kepler} dataset.

In Section \ref{sec:network_design} we describe our methodology for construction of the training set and the CNN architecture itself. In Section \ref{sec:application} we apply the network ensemble to the set of confirmed and candidate exoplanets in the \kepler\ data with orbital periods $>$ 10 days. In Section \ref{sec:discussion} we discuss the performance of the CNN and the ongoing challenges for this work, and we conclude in Section \ref{sec:conclusions}.

\begin{comment}
Hello astro-ph-leaks @LeaksPh ! Love your work. Follow me at @alexteachey 
\end{comment}
% Hello astro-ph-leaks @LeaksPh ! Love your work. Follow me at @alexteachey 

\section{Neural Network Implementation}
\label{sec:network_design}

\subsection{Motivation}
The goal of this work was to produce a fast and flexible methodology for identifying systems worthy of closer examination in the search for exomoons. To that end, we elected to utilize CNNs, which have been quite successful in searching for exoplanet transits and distinguishing them from false positives \citep[e.g.][]{shallue:2018, pearson:2018, zucker:2018, dattilo:2019,yu:2019, osborn:2020}. 

Ordinary, ``fully-connected'' artificial neural networks work by taking a series of inputs, multiplying all these values by some weights (iteratively derived to optimize performance) in one or more \textit{hidden layers}, and producing an output, which may be a classification or a regression. The structure of these networks can be simple or complex, but what these networks have in common is that these inputs (which may or may not be correlated in fact) are effectively independent features fed into the system. 

By contrast, a \textit{convolutional} neural network takes advantage of the fact that features in a given dataset (often an image) are not necessarily independent, but may be contributing to some structure in the data. For example, the presence of many red pixels scattered randomly about an image would tell you virtually nothing about what the picture represents; but a collection of them grouped together suggests you are seeing a red object, maybe say, a stop sign. The CNN then is designed to identify structures in a dataset by grouping together inputs, generating feature maps through convolution, reducing the dimensionality through \textit{pooling}, and ultimately learning to identify the relevant structures for the feature of interest.

In the present case, we wish to apply the CNN architecture to the problem of identifying exomoon signals in time-domain photometry. As in image processing, in the case of a light curve we do not wish to be distracted by the behavior of any individual data point; rather, we want to identify real structures in the data. For a transiting planet or moon, the associated flux reductions will be structured, and this structure is well known, so we may train a network to search for these features. In the meantime, there are other patterns in the data (for example, stellar variability), which are also structured, and therefore a CNN can in theory learn to discern and distinguish these patterns, as well.

For maximum flexibility, we wanted the CNN to take as its input a light curve segment containing a single planetary transit. This approach necessarily sacrifices encoded transit timing and duration variation information, which is typically an important component of identifying and validating candidate exomoons \citep[e.g.][]{TK18}. However, there are several motivations for opting for a single-transit approach. 

First, the CNN simply learns to identify moon transits based on transit morphology alone and therefore simplifies the problem considerably, since the classifier need not know anything about other transits in the light curve, nor do timings need to be known precisely ahead of time. Second, training on single events provides maximum flexibility for the classifier, allowing any planet to be analyzed regardless of the number of transits present. Third, it is trivial to apply a single transit classifier to any number of planet transits in a given system, and so if the CNN ensemble determines that multiple transits of the same planet contain moon-like signals, this can be an especially powerful indicator that the system deserves more scrutiny. Fourth, the advent of survey missions such as K2 and TESS, with more limited temporal baselines, means that many planets discovered in the future will have only a single transit recorded, or perhaps only a few. Finally, recent work \citep{hippke:2015, teachey:2018} appears to corroborate theoretical studies \citep{namouni:2010, spalding:2016} suggesting that moons are more likely to be found at a greater distances from their host stars. These planets at larger semi-major axes will therefore transit much more rarely and are likely to have one or only a few transits in existing datasets.

\subsection{Network training set}
Until it has been trained, a neural network is merely an algorithm. The network weights, derived through an iterative learning process, are what give the neural network its power. Therefore, the first step of producing a neural network is of course to start with constructing the training set; we will describe the CNN architecture in the following section.

Our first step for producing training and validation datasets was to generate noise-free light curves (transit models), which would subsequently be injected into real \textit{Kepler} light curves to match real observations as closely as possible. Because we currently lack a body of exomoon detections in the literature, these light curves must be generated artificially, but should be as physically plausible as possible and represent the full range of possible system architectures and geometries. We are ultimately pursuing a binary classification -- ``moon'' or ``no moon'' -- so we produce an equal number of light curves with and without moons. A balanced training sample is typically recommended for a binary classification problem such as ours, and indeed during experimentation we found that it was important to have equal representation of the two classes.

The noise-free light curves were made using the \LUNA\ code \citep{LUNA}, which produces a photodynamical model of the planet-moon system through nested-two body integration and modeling of 3-body syzygies. Each model consisted of a planet or planet-moon system with the planet's time of transit minimum centered midway through the time series. Each light curve length is fixed at 10 days, or 5 days on each side of the planet's time of transit minimum, since the CNN requires inputs of uniform size. We considered setting the window size to that of the Hill sphere (the maximum distance at which a moon could be found) but opted against this approach because 1) reliable masses are only available for a fraction of transiting planets, so reliable Hill sphere estimates are frequently unavailable, and 2) the light curves would have to be binned, therefore making the time sampling heterogeneous across the training sample. Moreover, binning is typically inadvisable for the moon search, as the signal may be lost in the process, so any re-scaling of the light curve for the sake of uniformity may be problematic. Meanwhile, the 10-day time window provides ample coverage to observe the entire Hill sphere for even very long period planets.

To produce a simulated light curve sample that matches closely the observed population of systems, we turned to the \textit{Kepler} sample itself for inputs. For each model light curve, a known planet-hosting star was selected at random from the list of KOIs on the NASA Exoplanet Archive. The relevant estimates for that star (mass, radius, density, surface gravity, effective temperature, metallicity, as well as the orbital period of the KOI) were extracted, and the appropriate limb darkening coefficients for that star were selected from a table of \kepler\ values calculated by \citet{sing:2010}. A simulated planet or planet+moon system was then generated, using these stellar properties as a frame of reference where appropriate. For example, the transit depth was dictated by the ratio of the planet size (randomly selected) to that of the star (drawn from the real distribution of \kepler\ stars). However, we do not draw the planet \textit{sizes} from these same KOIs, as that would severely limit the number of possible systems that could be simulated. The consequence of this is that we may in some cases be generating systems that are physically allowed but rare in nature, for example, low-mass stars with very massive planets. Still, it is worthwhile to generate training and validation sets that encompass the full range of possibilities, including systems we may be unlikely to find (or have not yet found) in nature.

There are a total of 14 inputs for \LUNA\ : the ratio of radii $p = R_P / R_{*}$, 
the stellar density $\rho_{*}$, 
the planet's impact parameter $b$, 
its orbital period $P_P$, 
two limb darkening coefficients $q_1$ and $q_2$ \citep[a non-standard reparameterization, see][]{ldcs}, 
the planet's density $\rho_P$, 
the semimajor axis of the moon $a_S$, 
a reference phase for the moon $\phi_S$, 
the moon's inclination $i_S$, 
longitude of the moon's ascending node $\Omega_S$, 
the ratio of radii for the moon and planet $R_S / R_P$, 
and the ratio of masses $M_S / M_P$. For the sake of producing the light curves and calculating a moon transit SNR -- and since TTVs are not incorporated into the learning process, we set the mass ratio to zero. The moon's orbit is circular in all cases ($e_S = 0$), which we generally expect for moons due to short circularization timescales.

\begin{figure}
    \centering
    \includegraphics[width=\columnwidth]{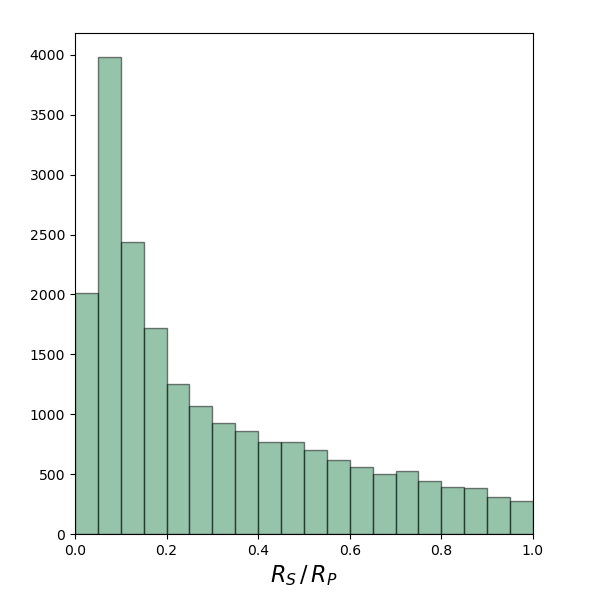}    
    \caption{Distribution of satellite-to-planet radius ratios in the training set.}
    \label{fig:RoR}
\end{figure}

The semi-major axis of the moon was selected randomly from a uniform distribution but was required to be somewhere between the Roche limit and $0.4895 \, R_{\mathrm{Hill}}$ for prograde moons and $0.9309 \, R_{\mathrm{Hill}}$ for retrograde moons \citep{domingos:2006}, based on the computed Hill sphere for the planet's mass and semi-major axis. The planet's period was chosen randomly from a log-uniform distribution between 10 and 1500 days; shorter period planets would not be considered. The orbital parameters for the injected moon were all drawn from a uniform distribution, as was the impact parameter of the planet. The moon's size was chosen from a uniform radius distribution between that of Europa and the Earth. Of course, smaller moons are likely lurking in the data, but if we have no hope of conclusively or convincingly demonstrating their presence with a full model fit, there is little point in training to detect them. Planet sizes were selected randomly from a uniform distribution between one half and 16 Earth radii. Figure \ref{fig:RoR} shows the distribution of satellite-to-planet radius ratios in the training set. Clearly, the training set includes some values for $R_S / R_P$ that are not found in the Solar System, but we wish to be agnostic about the limits of what may be found in nature.

The ``donor'' \kepler\ light curves, into which the noise-free transit models would be injected, were randomly selected from stars for which no planets have been reported. From these light curves a 10-day segment was chosen at random, our only requirement being that it lacked large data gaps, and into this light curve a simulated planet or planet+moon transit model could be injected. Of course, the fact that a planet has not been reported for the donor light curve does not necessarily mean there are no planets in the system, and particularly small transiting planets that have gone undetected could certainly mimic a moon signal if it appears near an injected planet. Identification of yet-undiscovered planets in these light curves is clearly beyond the scope of this work, and so we proceeded with the expectation that a small minority of our light curves designated as ``planet only'' could have additional signals embedded in them that could potentially confuse the CNN.

We also point out that the spectral type of the \kepler\ donor light curve selected for injection was not required to match the spectral type of the star used to produce the light curve model, and therefore the ratio of radii $R_P / R_{*}$ of the injected model could be unphysical for the true size of the ``donor'' \kepler\ light curve. For example, if the donor \kepler\ light curve belonged to a particularly large star, the transit depth of the injected model might suggest the transiter is a low-mass star rather than a planet. To the extent that photometric variability is correlated with spectral type, this could produce a light curve that is once again unlikely to be found in nature. The limb darkening coefficients for the injected transit, likewise, may not match the spectral type of the donor light curve. In any case, since we detrend the light curves prior to training, the astrophysical signatures of the donor star should in theory be effectively removed and therefore any discrepancy between the selected star and the modeled system should have also been removed. Furthermore, we do not train the CNN on any system parameters / observables, only the light curve itself, so the CNN has no knowledge about the star apart from the noise profile that it sees, the transit depths, and sometimes residual short-duration variability, which we attempt to screen as described below.

The light curves were detrended using the Cosine Filtering Autocorrelation Minimization (\cofiam) algorithm \citep{HEK2}, which optimizes a superposition of sinusoids for the purpose of removing astrophysical variability, while preserving short-duration signals (that is, those that may be produced by the transit of the planet or a moon). The planet's transit is also masked during detrending to ensure that the algorithm does not attempt to fit out the transit signal. We briefly experimented with employing a method-marginalized detrending approach as employed in \citet{TK18}, but ultimately decided against this as it was more prone to individual failures of one or more methods, and therefore could not be relied upon for a systematic approach without individual inspection of the detrended light curves.

Despite the robustness of \cofiam, stars with variability on short timescales (i.e. shorter than the transit duration) can slip through with detrended light curves showing residual variability that are too pronounced for our purposes, as they are capable of mimicking moon transits. Because we produced tens of thousands of simulated light curves, it was not feasible to examine each light curve by eye. Instead, we elected to screen poorly-detrended light curves by comparing the median absolute deviation (MAD) to the expected standard deviation of a clean light curve (a flat out-of-transit baseline), assuming the photometric noise is Gaussian. For a Gaussian distribution MAD is equal to $1.4826 \sigma$, but we point out that for a finite set of measurements there is an expected distribution of measured standard deviations, given by $\sigma_{sample} = \sigma / \sqrt{2(n - 1)} $, where $n$ is the number of data points. After some experimentation we elected to reject detrended light curves for which MAD $> 0.6745 \, (\sigma + 10\sigma_{sample})$. This does an excellent job rejecting most poorly-detrended light curves while also rejecting a small number of light curves which a by-eye treatment might have allowed for inclusion.

After screening, a total of 26,912 generated light curve segments were used in the training, amounting to 80\% of the total number of light curves utilized in training and validation. 10\% of the light curves (3364) were held back for validation during training as a check against over-fitting, and a further 10\% were held back for additional analysis of the results.

\begin{figure}
    \centering
    \includegraphics[width=\columnwidth]{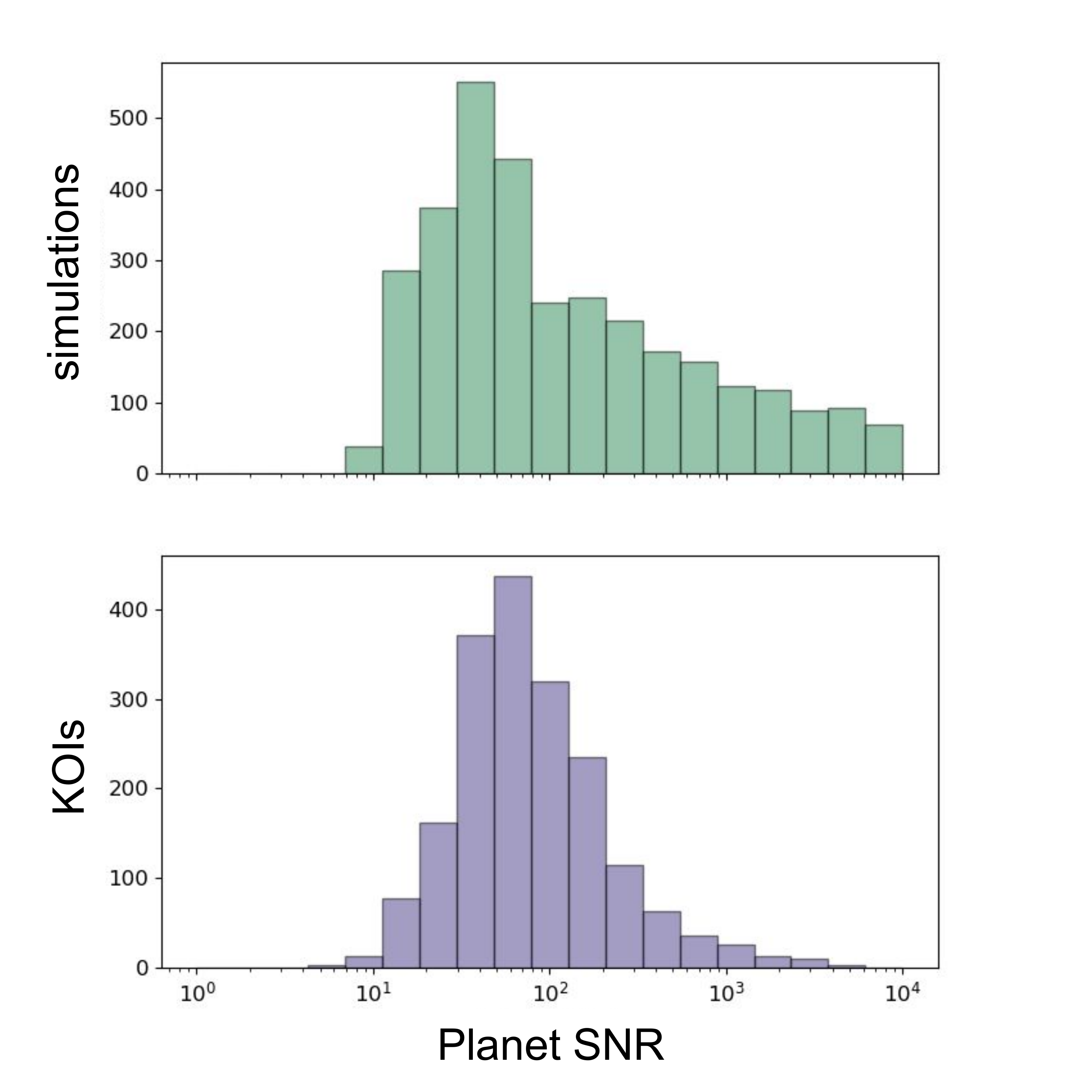}
    \caption{Comparison of the planet SNRs from the simulations (validation sample) and the KOIs vetted in this work. Moon SNRs are of course not available from the KOIs to compare with the simulations.}
    \label{fig:SNR_comparison}
\end{figure}

Each light curve was normalized for input as

\begin{equation}
    \boldsymbol{F}_{\mathrm{norm}} = \frac{\boldsymbol{F} - \mathrm{min}(\boldsymbol{F})}{\tilde{\boldsymbol{F}} - \mathrm{min}(\boldsymbol{F})} - 1,
\end{equation}

\noindent where $\boldsymbol{F}$ is the array of fluxes and $\tilde{\boldsymbol{F}}$ is the median value of the input segment. Because moons on high inclinations are not always guaranteed to transit, we ensured that any injected model that contained a moon but did not show a moon transit were reclassified as ``planet only.'' This results in having far more planet-only simulations on hand that planet+moon simulations, but training and validation was always balanced with equal number of each type. We also ensured that the planet SNRs are in good agreement with the distribution of SNRs found in the real \kepler\ data (Figure \ref{fig:SNR_comparison}). The true distribution of moon sizes in nature, and thus their associated moon SNRs, are currently unknown, and therefore cannot be used to ensure the training and validation sets are in good agreement with the real sample.

\subsection{CNN design}
Our CNN was constructed using the Python deep learning library \keras\ \citep{keras}, utilizing the \tensorflow\ backend \citep{tensorflow}. We initially adopted the architecture and hyperparameters of the CNN from those described in \citet{shallue:2018}. However, early testing suggested that this exact architecture was not able to produce the desired classification accuracy.

We experimented with a variety of hyperparameters. As each training run can take up to half an hour on a desktop machine, it is computationally intractable to explore every possible hyperparameter combination without relying on a high performance computing cluster (as employed by \citealt{shallue:2018}). We elected instead to explore hyperparameter space with an MCMC-type walker, initializing with values known to achieve decent results through experimentation.

Hyperparameter variables explored by the walker were 
1) the number of filters, 
2) kernel size, 
3) pool size, 
4) pool type (average or max pooling), 
5) stride length, 
6) dropout rate, 
7) the number of convolutional layers, and 
8) the number of dense layers. We refer readers to the \keras\ documentation for details on these options.

\begin{figure}
    \centering
    \includegraphics[width=0.5\columnwidth]{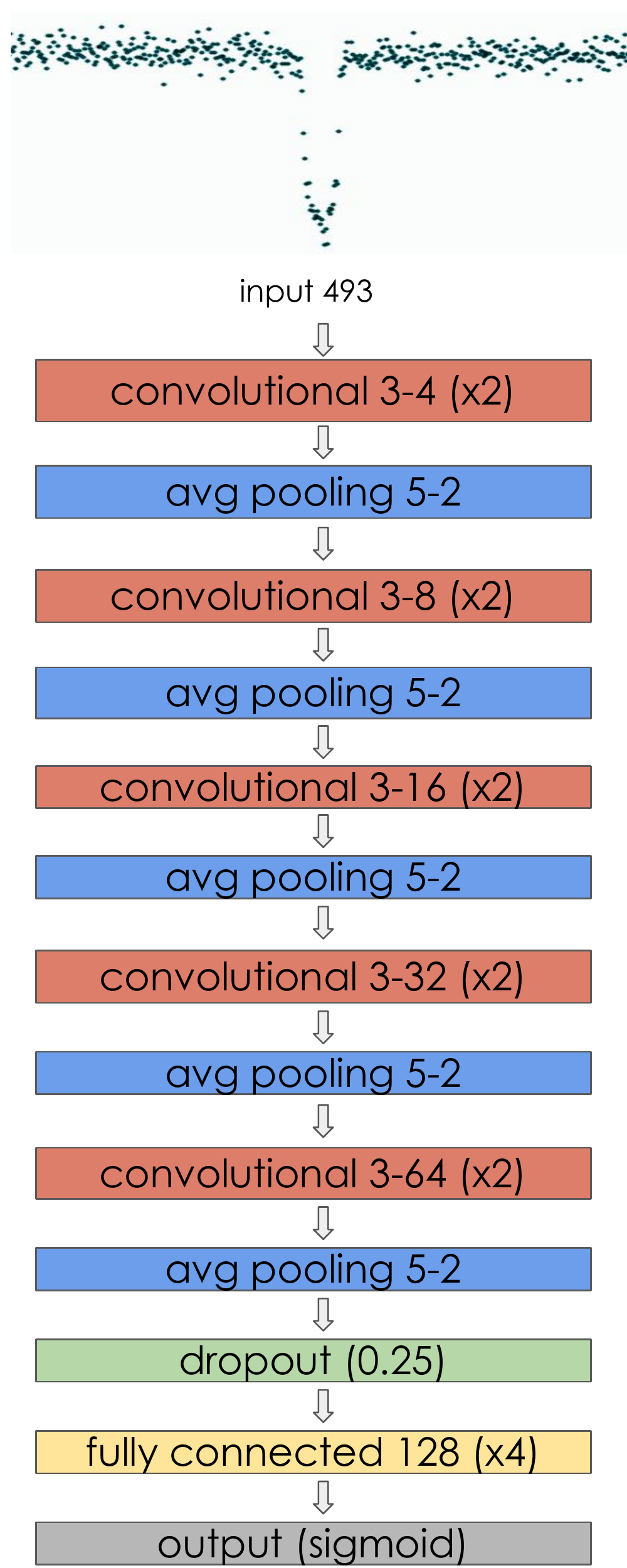}
    \caption{Example architecture one convolutional neural network in the 
    ensemble. Following \citealt{shallue:2018}, the convolutional layers are 
    marked as \textit{kernel size - \# of filters}, while the pooling layers 
    are marked as \textit{pool size - stride length}. Each convolutional 
    layers is actually two layers in sequence, before each pooling (which may 
    be average of max pooling). All hyperparameters indicated here were variable.}
    \label{fig:CNN_architecture}
\end{figure}

A sample architecture can be seen in Figure \ref{fig:CNN_architecture}. The kernel initializer was orthogonal, and all activations were \texttt{relu} (rectified linear unit), except for the final activation which used \texttt{sigmoid}. The loss function was set to ``categorical crossentropy''.  The optimizer was ``adam'' \citep{kingma:2014}, and we optimized for classification accuracy. Experimentation with these choices showed no meaningful impact on the network performance. We used balanced-batch sampling, and each architecture was trained until validation accuracy had failed to improve after five epochs. We found that the last improved epoch was generally the point beyond which the model started overfitting. 

\begin{figure*}
    \centering
    \includegraphics[width=2\columnwidth]{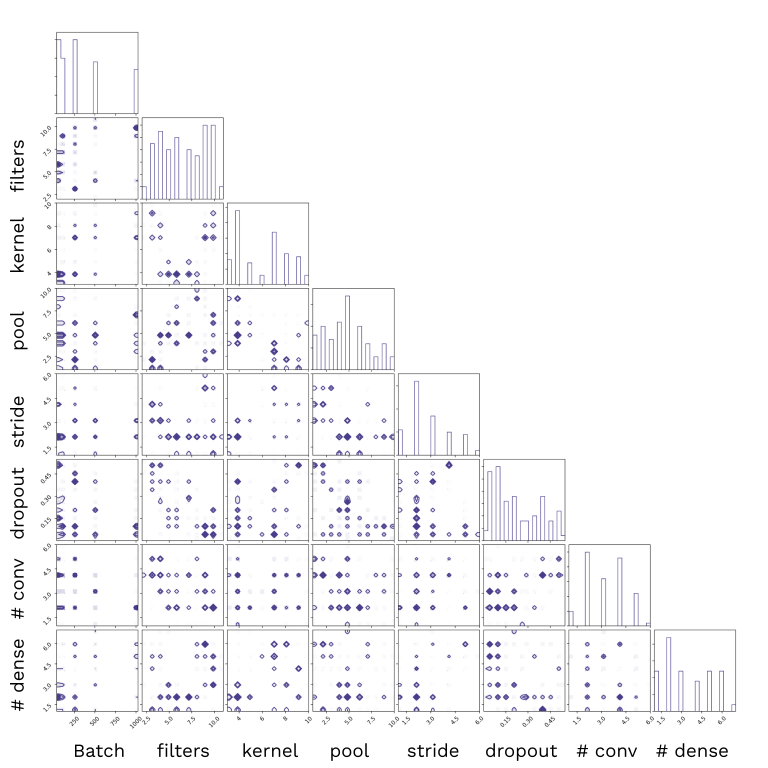}
    \caption{Distribution of hyperparameters for every CNN with validation 
    accuracy $> 80\%$.}
    \label{fig:hyperparameter_corner}
\end{figure*}

We found that in fact a variety of different hyperparameter combinations produced very similar validation accuracies, though there is no apparent pattern that points towards an optimum hyperparameter combination. (Similarly, there was no apparent pattern to the architectures that performed poorly). Importantly, trained CNNs with effectively identical classification accuracies did not necessarily make the same predictions for a given transit. This suggested that it might be possible to leverage the power of an \textit{ensemble} of classifiers to boost precision, analogous to the production of a random forest through an ensemble of decision trees. Altogether we used the best 50 CNN classifiers, all with individual classification accuracies $> 80\%$. The distribution of hyperparameters for every CNN achieving $> 80\%$ accuracy can be seen in Figure \ref{fig:hyperparameter_corner}. It is apparent that there is no obvious pattern to hyperparameter combinations that produced good results.

For our purposes, we are chiefly interested in identifying the most promising moon candidates. Computing population statistics are beyond the scope of this work; if that were our goal, we would be considerably more concerned with understanding the underlying performance of the CNN classifiers. Given our aim, we are comparatively less concerned by discarding those systems that may have moons present, but which the networks either fail to identify them or do not have sufficient confidence in a ``moon''. This is to say, we wish to minimize the false positive rate and maximize \textit{precision}, defined as the ratio of true moon classifications over the sum of all moon classifications (true and false positives). This comes at the expense of \textit{recall}, defined as the ratio of true moon classifications to the total number of moons in the sample (true positives and false negatives).

Every viable transit -- that is, those with a 5-day baseline present on either side of the planet's transit -- would be analyzed for every planet (see the next section for the results of this analysis). For each transit, a classification and a probability for that classification (from the sigmoid function) would be produced for every voting model. The probabilities are then used to augment the binary classification (1 for a moon, 0 for no moon) through a weighted average. That is:

\begin{equation}
\centering
    \overline{x} = \frac{\sum_i^n w_i x_i}{\sum_i^n w_i},
\end{equation}

\noindent where $\overline{x}$ is the final classification of the transit in question ($>0.5$ for a moon, $<0.5$ for no moon).  The classification is further modified by using the ``agreement metric'', which allows us to tune precision and recall as desired. The agreement metric is defined such that complete agreement of the voting models has a value of 1, and maximum disagreement (a 50/50 split) has a value of zero. As shown in Figure \ref{fig:precision_recall}, we can achieve higher confidence in our moon predictions when we require the models to be in good agreement (setting the agreement metric close to 1). Thus, a final moon prediction is made only in cases where the agreement metric surpasses our set threshold, otherwise it is re-classified as a no-moon prediction.

Unlike \citet{shallue:2018} and follow-on papers, we opt for using a single light curve input, rather than two parallel light curve inputs. In that work, the authors sought to distinguish genuine planets from false positive scenarios, and were therefore interested in the information that extensive out-of-transit data might reveal (for example, a phase curve and/or a secondary transit) as well as the morphology of a zoomed-in transit. Both of these input formats required binning of the light curve in order to make the input sizes uniform across a range of planet periods. In our case, however, binning is an inappropriate choice for the moon search, as it has the potential to wash out moon transits, and at the same time, there is little to be gained by employing a ``global'' view, since the moon will necessarily be located in close physical and temporal proximity to the host planet.

\subsection{Network validation}
\subsubsection{CNN training results}
As previously mentioned, many individual CNN architectures were able to achieve up $\sim 80\%$ accuracy. Many other hyperparameter combinations predicted no better than random. While $80\%$ classification is not terrible, and is even acceptable in some contexts, it was unsatisfactory for our purposes.

\begin{figure}
    \centering
    \includegraphics[width=\columnwidth]{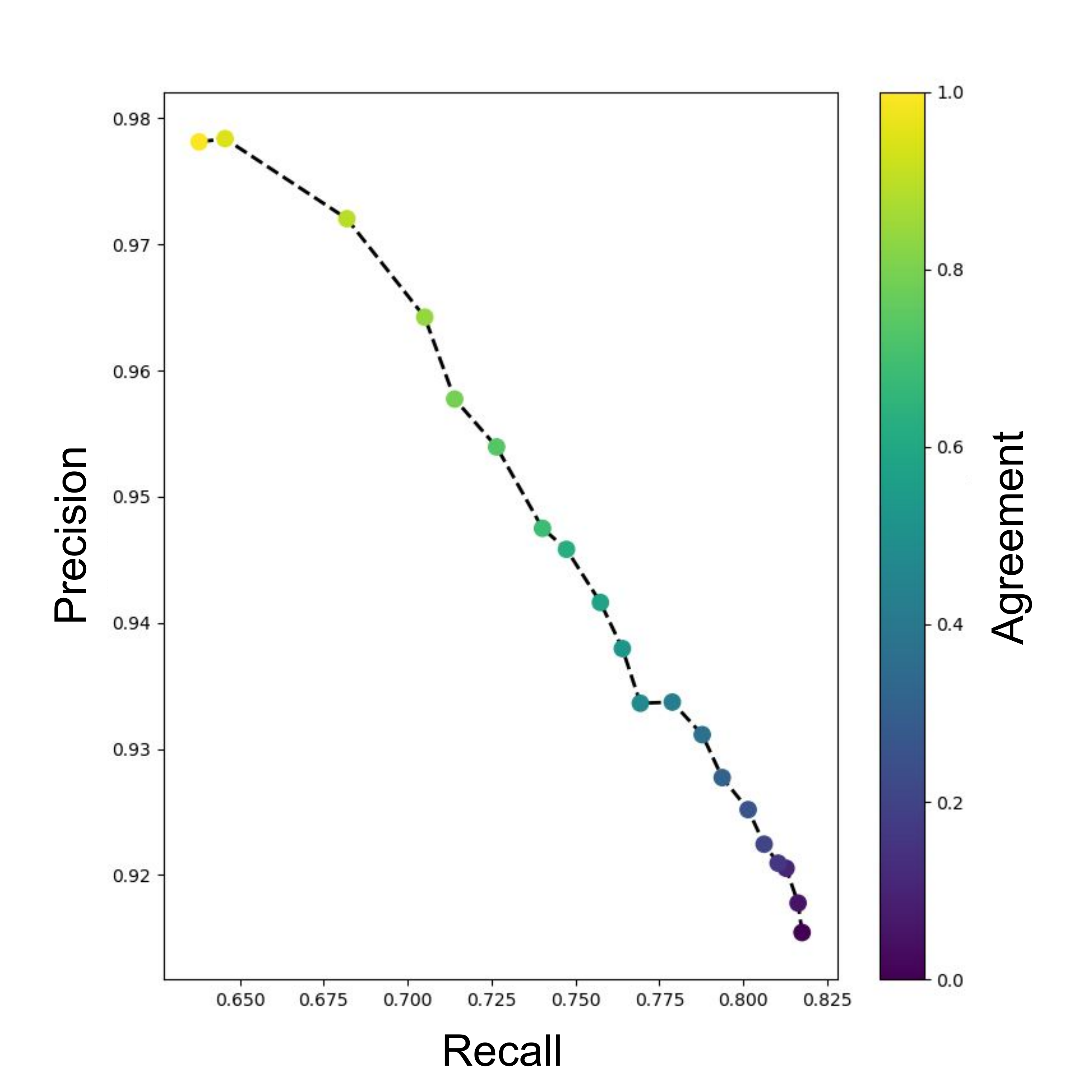}
    \caption{Precision-recall plot of the CNN ensemble on (simulated) validation data, 
    indicating that maximum precision is acquired when the ensemble is also in total agreement.}
    \label{fig:precision_recall}
\end{figure}

As shown in Figure \ref{fig:precision_recall}, the results are far better when we leverage the power of a network ensemble. We are able to achieve precision of up to 97\% when the agreement metric for the CNN ensemble = 1. Once again, we define the agreement metric such that 1 corresponds to all CNN networks in agreement, and 0 means maximum disagreement (the number of ``moon'' classifications is equal to the number of ``no moon'' classifications). We note that even in the cases of large disagreement in the classifications, precision remains in the range of 90\%, and recall is substantially higher.

\begin{figure}
    \centering
    \includegraphics[width=\columnwidth]{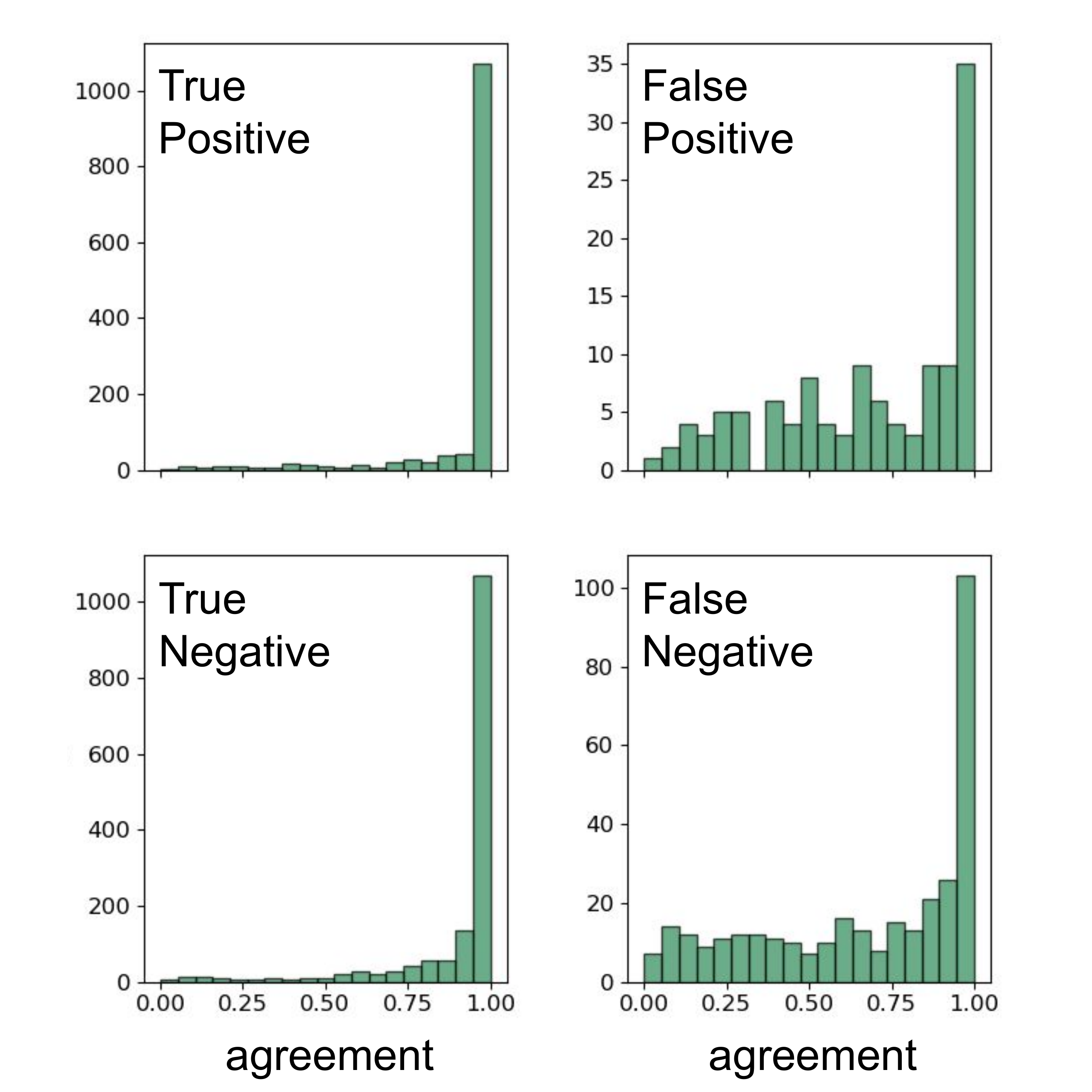}
    \caption{Distribution of true/false positives/negatives as a function of ensemble 
    agreement for the validation light curves (simulated data). The majority of 
    classifications come with maximum ensemble agreement. Note the differing scales.}
    \label{fig:TFPN_histograms}
\end{figure}

Figure \ref{fig:TFPN_histograms}, meanwhile, shows the distribution of true positive (TP), false positive (FP), true negative (TN), and false negative (FN) classifications during the validation process, as a function of agreement metric. Encouragingly, the vast majority of classifications are true positives and true negatives, and most of those cluster around the highest agreement metric. By contrast, false positives and false negatives are a bit more uniformly distributed across all network agreements (though still peaking strongly at high agreement metric), and fortunately represent a small share of the validation set. Following these numbers, we may be reasonably confident in both the classification accuracy and the utility of the agreement metric.

\begin{figure}
    \centering
    \includegraphics[width=\columnwidth]{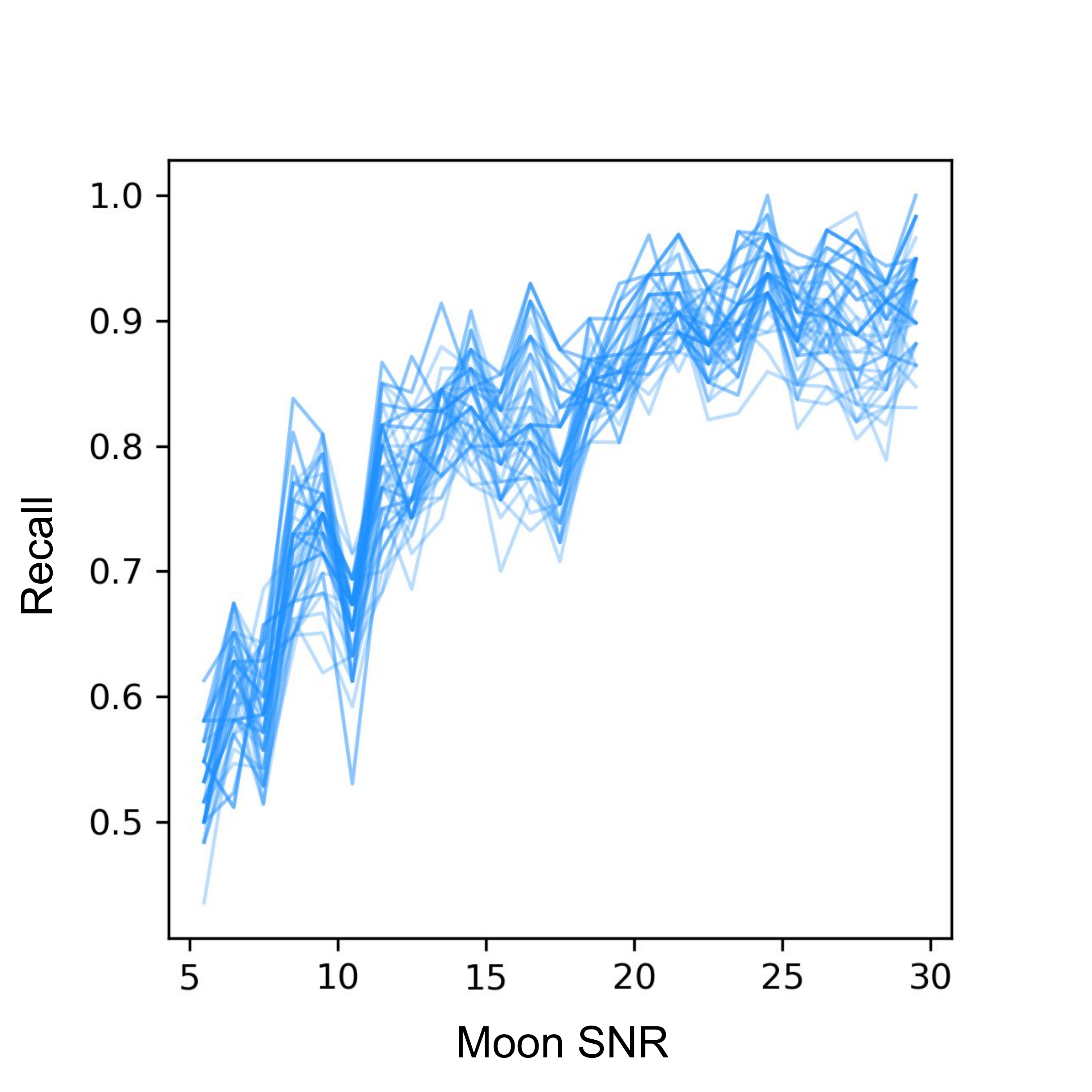}
    \caption{Recall (the fraction of moons that are successfully recovered) as a function of moon SNR.}
    \label{fig:recall_vs_moonSNR}
\end{figure}

It is also worth understanding the network's performance as a function of moon transit SNR. Figure \ref{fig:recall_vs_moonSNR} shows recall -- the fraction of light curves containing moons that are correctly identified as such -- as function of moon SNR for all 50 voting models. We note that it is not possible to plot precision or accuracy as a function of moon SNR, as these calculations incorporate the number of false positives and true negatives, respectively; clearly there can be no moon SNR for light curves that lack a moon entirely. Unsurprisingly, the fraction of moons that are successfully identified in the validation set climbs as moon SNR increases. The spread in recall values at each moon SNR bin gives some indication of the diversity of voting models which collectively improve classification accuracy.

\subsubsection{False-positive test light curves}
In addition to the CNN training statistics above, we desired an additional test for classification accuracy, which could quantify a false positive probability on a star-by-star basis. We discuss this in greater depth in Section \ref{sec:application}. Because each light curve presents unique astrophysical variation, which may have been preserved even after detrending and screening, there is the potential that these systematics may be capable of fooling the CNN ensemble into an erroneous moon classification. To guard against this scenario, we produced a false positive test sample for each individual KOI, consisting of at least 100 planet-only transits for each target to also be vetted by the CNN ensemble.

Following the neighbor-masking procedure described later (Section \ref{sec:data_prep}), we masked every transit present in the star's light curve, including the target planet and any of its neighbors. We then injected planet transits at random locations throughout the light curve, adopting the fiducial parameters of the target planet from the NASA Exoplanet Archive.  These light curves were then detrended and segmented in the same way as the real \kepler\ light curves, and fed through the same CNN ensemble classification pipeline to produce a false positive rate for each KOI. That is, since every injected transit is known to only contain a planet transit (no moon), moon classifications will all be erroneous, and we therefore have a metric for how often the astrophysical variation of any given star in the sample produces a false positive.

\subsubsection{Feature mapping}
\begin{comment}
\begin{figure*}
    \centering
    \includegraphics[width=17cm]{figures/simulation_feature_maps.pdf}
    \caption{Feature maps for six example simulations containing moon signals. Red data points indicate locations where the moon prediction is unchanged by replacing the light curve with a median filter. The moon prediction changes to a no moon prediction when the blue points are replaced by a median filter.}
    \label{fig:sim_feature_maps}
\end{figure*}
\end{comment}

\begin{figure}
    \centering
    \includegraphics[width=\columnwidth]{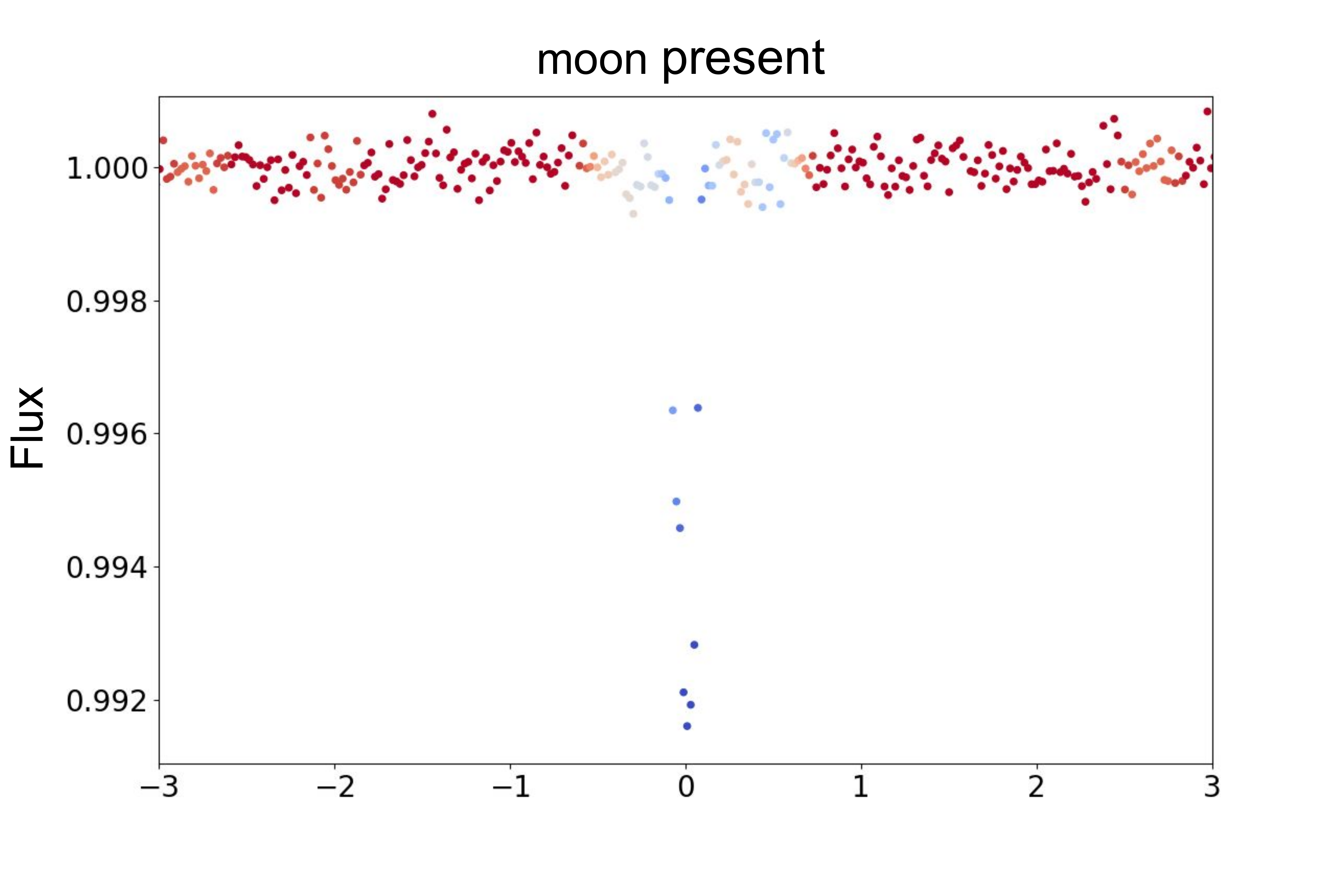}
    \includegraphics[width=\columnwidth]{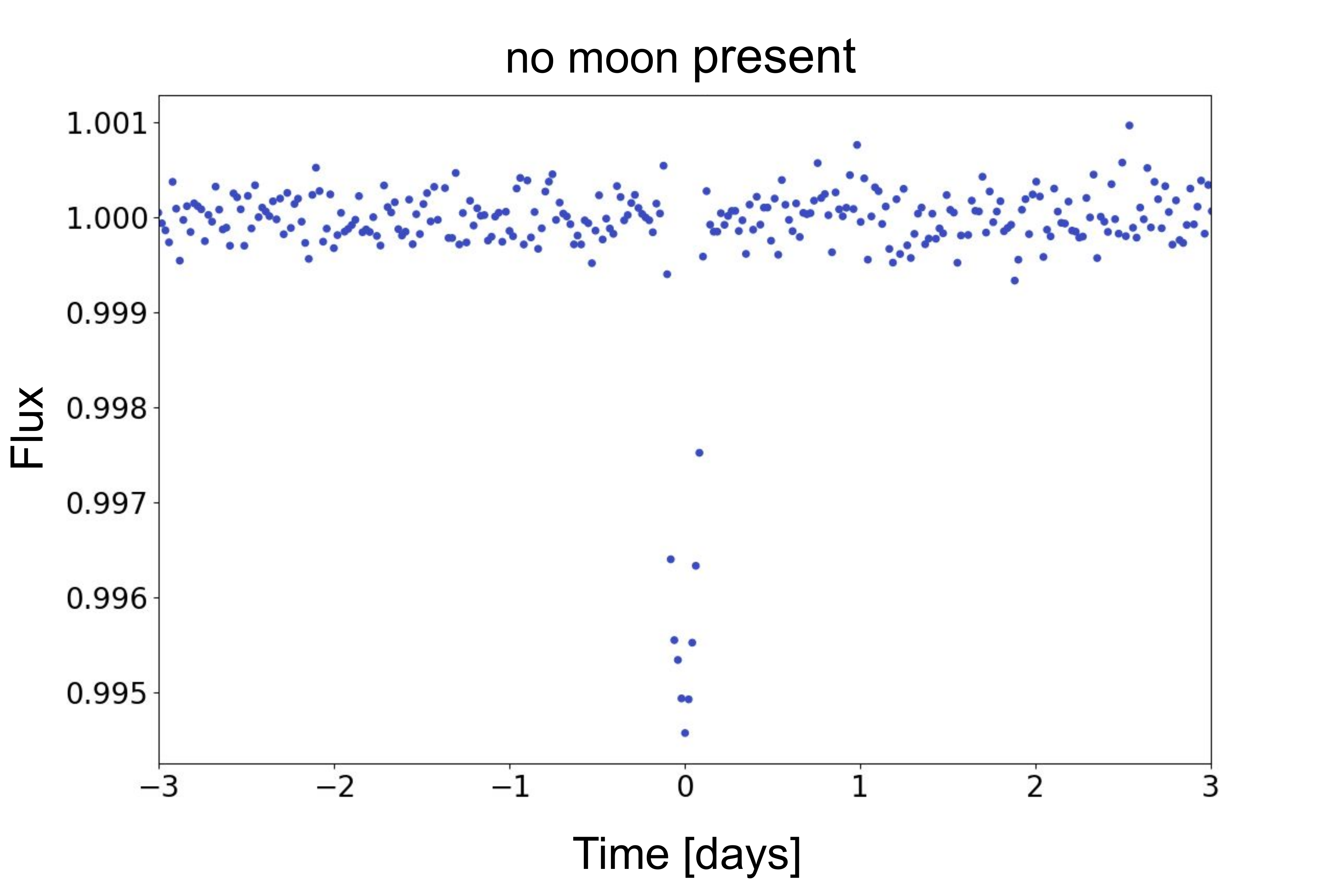}
    \caption{\textit{Top:} Feature map of an example simulated light curve containing a moon, indicating the region where masking results in a ``no moon'' prediction (in blue). In this and most cases, the morphology of the planet transit is responsible for the moon prediction. \textit{Bottom:} a comparable light curve without a moon, where the planet transit is not triggering a moon classification.}
    \label{fig:sim_feature_maps}
\end{figure}

It is worthwhile to consider what features the CNN ensemble is actually ``seeing'' when it makes a moon classification. To this end we can produce a feature map, highlighting regions of the light curve that are responsible for a moon prediction. This test was performed in, for example, \citealt{shallue:2018}, who demonstrated that their CNN identifies secondary transits as a key discriminator between genuine planets and eclipsing binaries. In our case, we wish to see whether moon transits in the light curve are what actually produces the moon prediction, or whether other light curve features (say, stellar activity) may be confounding the detection. The training process makes clear that we are doing well in detecting moons, but a visual aid is nevertheless helpful.

To this end, we employed a moving window replacement of fluxes to examine where the prediction for the light curve changes. In so doing, we can identify the regions that are responsible for the classification. We utilized a median filter with a width equal to four times the planet duration, and replace segments with a width equal to two transit durations, using Gaussian noise around the median filter line based on the photometric errors. This removes any features of comparable duration to a planet and moon transit, and then we ran these modified light curves through the CNN ensemble as before. We initialized an array of zeros with length equal to that of the light curve, and added or subtracted one at each index in the feature replacement window if the prediction was ``moon'' or ``no moon'', respectively. These values can then be used to produce the color map indicating the data points that are responsible for one prediction or the other.

Example results of this test can be seen in Figure \ref{fig:sim_feature_maps}. Red points indicate regions where the prediction of the light curve with this segment replaced result in a moon classification, while blue corresponds to regions where masking results in a no moon classification. As such, for light curves containing a moon, blue highlights the portion of the light curve that is responsible for the moon prediction; removing this segment will result in a no moon classification. Ideally, we want to see blue regions in the moon-containing light curves that are well localized around moon features. For light curves that do not contain a moon, we want to see a light curve that is entirely blue.

Unsurprisingly, we find that the classifier performs well in most cases, but can be confused by particularly noisy or poorly-detrended light curves. We find that the planet's transit morphology is predominantly what the CNN uses to distinguish systems with a moon from those without. Typically this means that the moon transit is coincident with that of the planet, resulting in a distinctively asymmetric dip. Detached moon transits, occurring far from the planet's transit, are sometimes identified, but are also frequently missed. This behavior may be due to an imbalance in the training set, whereby moon transits are more often than not seen at or near the planet transit. Or, it could be that detached moon transits are generally more difficult to discern from stellar activity. Yet another possibility is that moon features in the vicinity of the planet's transit preferentially survive detrending by virtue of the transit masking. In any case, this test indicates that the CNN will be more sensitive to close-in moons or those seen near conjunction. Plausible moon classifications will therefore likely resemble the top of Figure \ref{fig:sim_feature_maps}.

\section{Application to the \textit{Kepler} Data}
\label{sec:application}

\subsection{Data preparation}
\label{sec:data_prep}
It is important that there are no major differences between the training set and the data to be classified, as these differences have the potential to undermine the classification accuracy obtained during the training process. Just as we have endeavored to build a realistic training set, then, we must also ensure that we process the real data in the same way we processed the artificial training set. 

To prepare the real \kepler\ data, we first pulled planet transit timings from \cite{holczer:2016} to extract the light curve with the identical 5-day window on either side of that transit. These light curves were then detrended using the same \cofiam\ algorithm as utilized on the simulated data, normalizing, and rejecting once again badly detrended light curves by evaluating the median absolute deviation.

A large fraction of KOIs are multiplanet systems. Systems containing more than one transiting planet present a complication for the moon classifier, as transits of a neighboring planet can and will mimic a genuine moon signal. Restricting our moon search to planets without neighbor contamination would severely limit the number of systems / transits that could be examined, and would rule out the examination of some planets entirely. Moreover, our training set contained no other nuisance planet transits. As such, we elected to remove contaminating transits of neighboring planets. 

\begin{figure}
    \centering
    \includegraphics[width=\columnwidth]{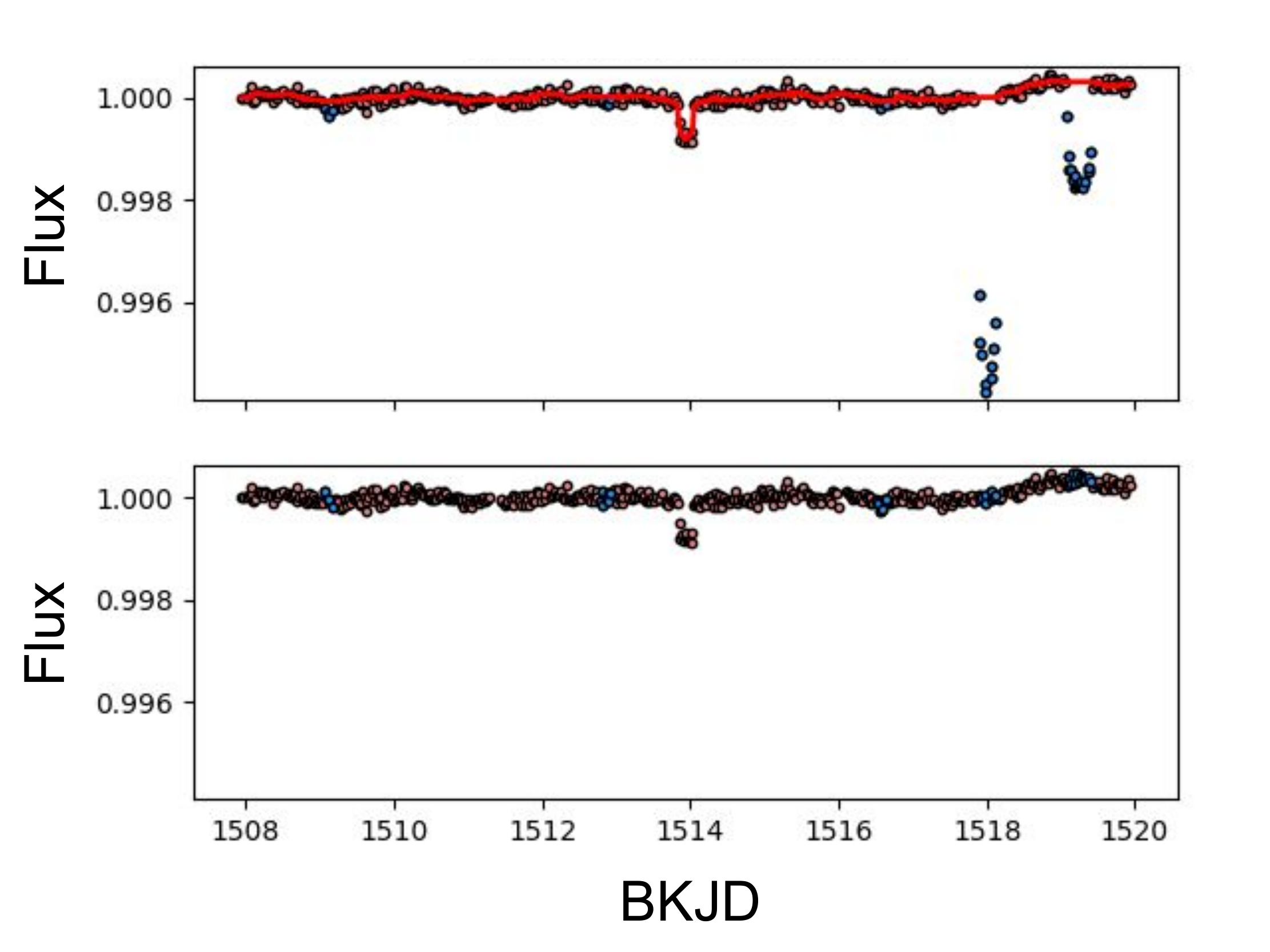}
    \caption{An example light curve demonstrating the removal of contaminating transits by neighboring planets in the system.
    These transits will mimic moon signals if they are not removed. The red line is the median filter, while transit locations are marked with blue points.}
    \label{fig:contam_replacement}
\end{figure}

To achieve this we computed a moving median trend line for the light curve, masking the neighbor's transit (also using transit timings from \citealt{holczer:2016}, where available), and generating Gaussian-noise around the trend based on the noise profile of the light curve (see Figure \ref{fig:contam_replacement}). While this approach has the potential to introduce a localized removal of any correlated noise, particularly variability with duration shorter than the neighbor's transit duration, we found this to be the most robust approach to the problem. Gaussian processes, by contrast, were significantly more computationally expensive, did not always produce a reasonable result, and were locally flat in the region of the masked transit, thus obviating the need for this more sophisticated approach. To the extent that published transit timings might be erroneous, we may not in fact be removing all the neighbor transits. The presence of these neighbors was always recorded, however, so that any potential contaminants could be taken into account when considering a predicted moon signal.

We initially produced 10-day segment light curves for 5400 KOIs, regardless of the planet's period and its NASA Exoplanet Archive disposition (confirmed, candidate, or false positive). This resulted in 1,190,008 transit segments to be vetted with the CNN ensemble. In our final analysis we vetted 1880 KOIs, comprised of 1036 that are classified as 'confirmed' and 844 classified as 'candidate' planets by the NASA Exoplanet Archive, and which had periods longer than 10 days. This amounted to 57,072 transit events.

\subsection{Application to real light curves}

Of the 57,072 transit events in the final sample, the CNN ensemble classified a mere 845 transits (roughly 1.5\%) as showing evidence for a moon. These events were associated with a total of 194 different planets, or about 10\% of the 1880 KOIs examined. Only 177 of these transits (0.3\%) met our highest standard of confidence, with a agreement metric greater than 0.9. This metric therefore provides additional discrimination power by about a factor of 5. The mean false positive rate across all vetted systems based on the planet-only light curve injection test was 0.03\% (median of zero). The full results of the CNN ensemble classification can be found at \url{https://github.com/alexteachey/KOI_CNN_moon_results}.

A total of 65 KOIs (3.5\%) were classified as having more than 50\% of their transits containing moons by a simple majority vote. Of these, 39 are considered ``confirmed'' planets by the NASA Exoplanet Archive, while 26 remain ``candidates''. Applying the agreement metric threshold, were are left with a mere 10 systems with a moon classification in $\geq$50\% of the transits, 6 of which are from confirmed planets and 4 from candidate planets. Altogether there are 89 transit events classified as having a moon and surpassing the agreement threshold. We further reject KOIs 423.01, and 3683.01, and 3943.01, as these systems had more transits rejected for failing the MAD test than were vetted by the CNN. It is therefore likely that the transits that were vetted are false positives by virtue of stellar activity mimicking a moon signal. We also reject KOI-94.01, for poor detrending and unmitigated contaminations from KOI-94.02 and KOI-94.03. Of the six remaining systems, two of them -- KOI-1032.01 and KOI-1192.01 -- are long period planet candidates with only a single analyzed transit. Clearly this is not much to work with, both for the CNN ensemble, and for subsequent follow-up. These two KOIs also show transits morphologies that suggest they could be eclipsing binaries. Were that to be the case, a moon-like transit in these light curves could in fact result from the presence of an S-type planet. 

Table \ref{tab:final_kois} presents relevant statistics for the systems which emerge as the most promising based on the cuts we have just described. However, we caution that these systems should not be considered \textit{exomoon candidates} as this time. They may, however, deserve further scrutiny in subsequent work.

Visual inspection of the feature maps for these KOIs (see section 2.4.3) reveals that the moon classifications for these systems, like the simulated light curves, are due to the planet (candidate) transit morphologies. Once again, blue points are associated with regions of the light curve that are responsible for a moon classification, removal of which changes the light curve classification to ``no moon''. We see no evidence of detached transits flagged by the CNN ensemble, nor do we see obvious moon-like features in these light curves. This is frequently the case for simulated systems containing small moons, however. It is also worth keeping in mind that moon transits are not always readily apparent in phase-folded transits, as the moon transit will appear in a different location each epoch; we present phase-folds here for want of a cleaner way to display so many individual transits. It is noteworthy that all of these KOIs present very clean light curves with relatively deep transits, and perhaps not surprising then that these systems are among the less ambiguous classifications made by the CNN ensemble.

\begin{table*}
    \renewcommand{\arraystretch}{1.3}
  \centering
  \begin{tabular}{|c|c|c|c|c|c|c|c|}

  \hline
    \textbf{KOI} & \textbf{Disposition} & \textbf{Period [d]} & \textbf{\# transits} & \textbf{\# MAD failures} & \textbf{\% moons} & \textbf{\% moons (agreement)} & \textbf{median agreement} \\
    \hline 
    \hline
KOI-192.01 & CONFIRMED & 10.29 & 98 & 0 & 90.82 & 50.0 & 93.33 \\
\hline
KOI-631.01 & CONFIRMED & 15.46 & 35 & 10 & 100.0 & 51.43 & 90.91 \\
\hline
KOI-686.01 & CANDIDATE & 52.51 & 17 & 1 & 94.12 & 70.59 & 100.0 \\
\hline
KOI-1032.01 & CANDIDATE & 1500.14 & 1 & 0 & 100.0 & 100.0 & 95.24 \\
\hline
KOI-1192.01 & CANDIDATE & 1295.36 & 1 & 0 & 100.0 & 100.0 & 94.87 \\
\hline
KOI-3680.01 & CONFIRMED & 141.24 & 6 & 0 & 100.0 & 83.33 & 96.97 \\
\hline

 \end{tabular}
  \caption{Table of systems identified by the CNN ensemble as most promising for containing exomoon transits. Disposition refers to the KOI's status as listed on the NASA Exoplanet Archive (confirmed or candidate planet). The number of transits refers to the number vetted by the CNN, excluding the number of transits rejected for failing the MAD test or any other requirement. the second to last column refers to the percentage of transits classified as having a moon and surpassing our agreement threshold (90\%). The final column reflects the median agreement across all transits vetted, whether they are predicted to have moons or not.}
  \label{tab:final_kois}
\end{table*}

\begin{figure*}
    \centering
    \includegraphics[width=16cm]{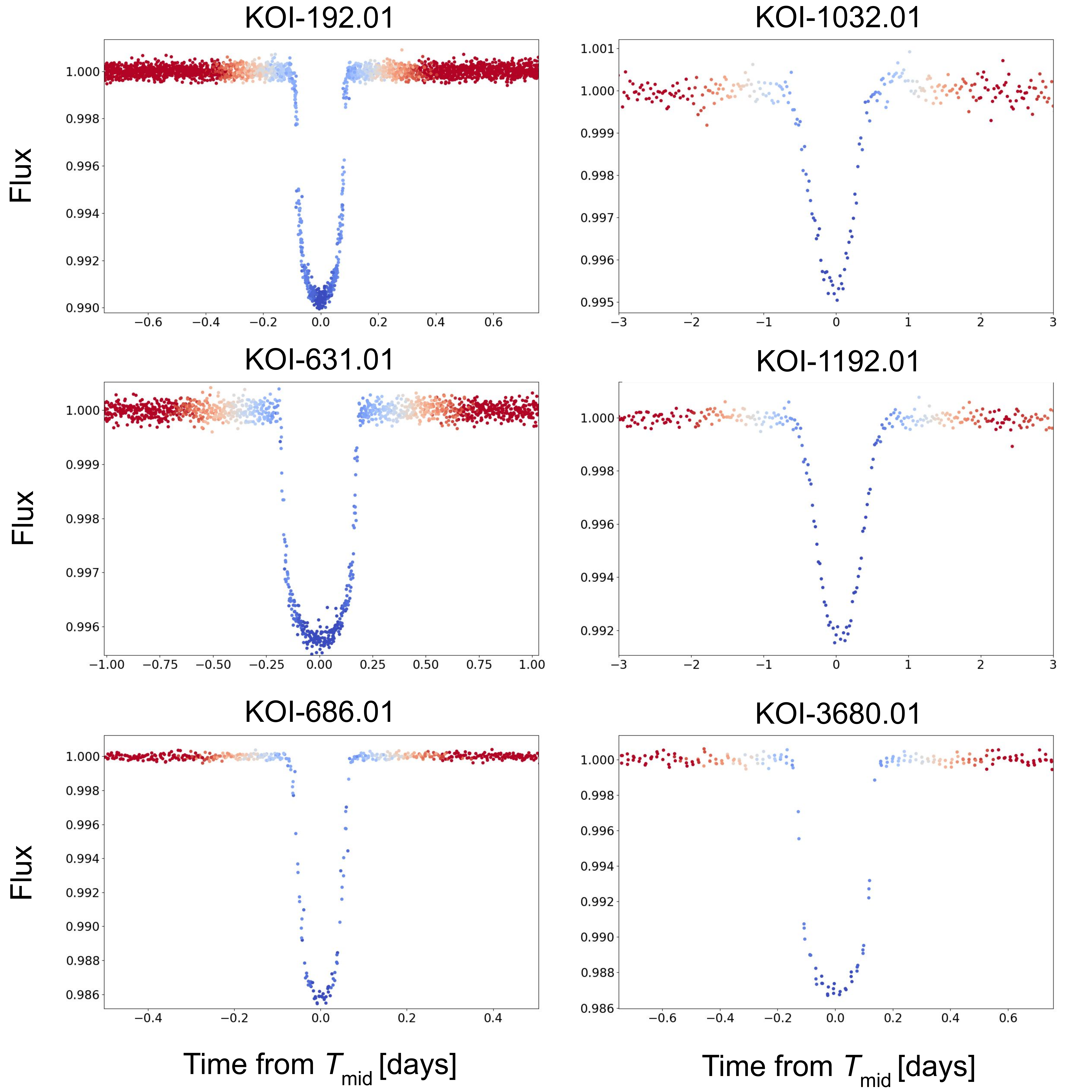}
    \caption{Phase-folded feature maps of the six KOIs presented in Table \ref{tab:final_kois}. Any TTVs present in these systems are removed to show transit morphology more cleanly. The methodology is as described in section 2.4.3; blue indicates areas of the light curve that are contributing to the moon classification, as their removal results in a ``no moon'' classification. KOIs 1032.01 and 1192.01 appear to have morphology consistent with eclipsing binaries, though as of this writing they remain listed as planet candidates on the NASA Exoplanet Archive.}
    \label{fig:final_koi_phasefolds}
\end{figure*}

\begin{figure}
    \centering
    \includegraphics[width=\columnwidth]{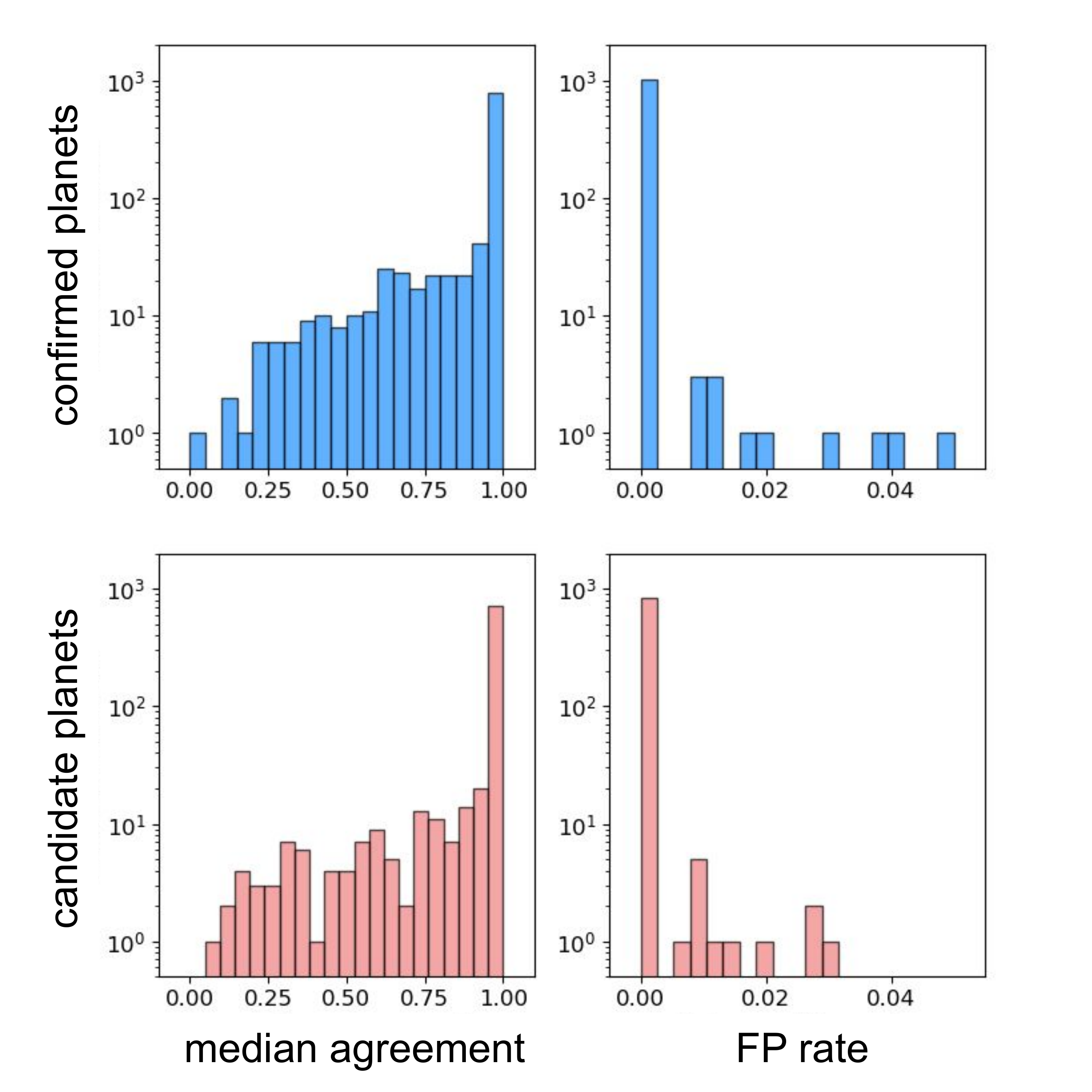}
    \caption{Distribution of median agreement and false positive rates for confirmed (top) and candidate (bottom) planet KOIs. Note the logarithmic scale of the $y$-axis.}
    \label{fig:agmet_and_FPrates}
\end{figure}

As we see in Figure \ref{fig:agmet_and_FPrates}, the real, full sample of KOIs shows broadly consistent median agreement metrics on average with the training sample (Figure \ref{fig:TFPN_histograms}), with the majority of cases showing an agreement metric near 1. False positive rates, computed from the individual false positive test light curves, are nearly all at or below 1\%.

\begin{figure}
    \centering
    \includegraphics[width=\columnwidth]{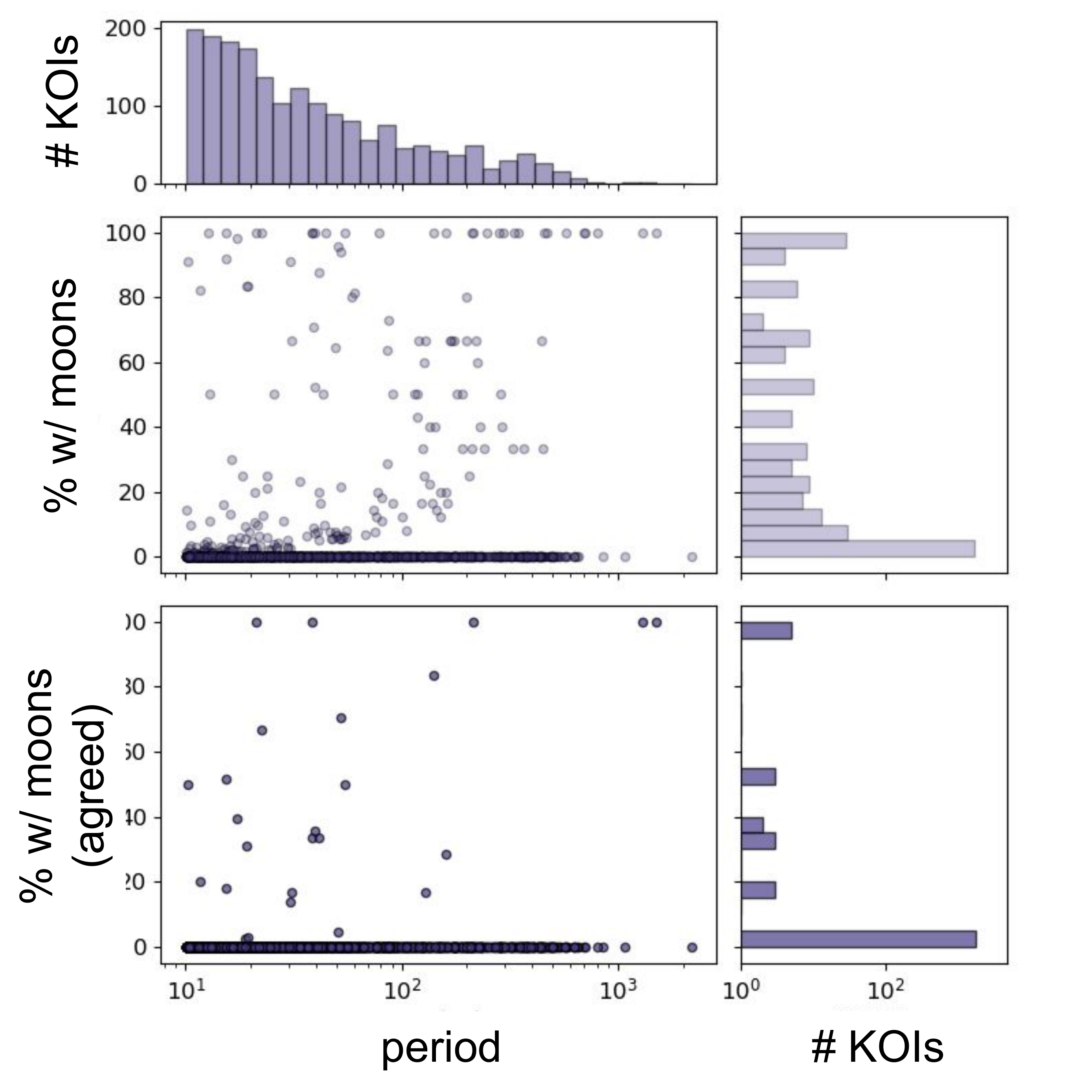}
    \caption{\textit{Top:} Distribution of KOI periods in the final sample. \textit{Middle:} The percentage of transits from each system classified as containing a moon, by simple majority vote / no agreement threshold applied. A histogram of these results is adjoined. \textit{Bottom:} The percentage of transits from each system classified as having a moon after applying an agreement metric of 0.9. The vast majority of systems. The vast majority (98\%) are classified as having $\leq$ 5\% of their transits containing a moon signal.} 
    \label{fig:frac_wmoons_vs_period}
\end{figure}

Despite the relatively high confidence we may have in the CNN's performance on simulated data, the ensemble's application to real data requires further scrutiny. Because the underlying exomoon population in the \kepler\ sample remains essentially unknown, it is difficult to know if the number of moon predictions the CNN ensemble produces is too high or too low. We note, however, that \citep{teachey:2018} provided some basis for comparison, having found a low occurrence rate $\eta = 0.16^{+0.13}_{-0.10}$ of Galilean analogs in the \kepler\ data.

When applied to the real data, the CNN ensemble returns a very small number of moon predictions to high confidence. This result may cut against our (perhaps naive) expectations that moons will be abundant in exoplanetary systems, at least those at some distance from their host star. Even so, this result might not be entirely unexpected, in view of the \textit{Kepler} photometry's precision, which is simply not high enough to detect small moons in many cases. In this sense, we should be careful about over-interpreting the present results as implying a dearth of moons. Rather, the results indicate that even a robust machine learning approach can fail to locate low SNR signals in this dataset.

\section{Discussion}
\label{sec:discussion}

\subsection{Applicability of the simulated sample}

The question invariably arises, to what extent can we trust the classifications made by the CNN? As detailed above, we have tried to address this concern in several ways: first, by producing a realistic simulated dataset, mimicking the distribution of real \textit{Kepler} planets and utilizing real light curves for model injection; second, by optimizing for classification accuracy during the training and validation process; third, by utilizing many, independent CNN architectures to vote on a classification, thereby boosting accuracy through classification agreement; and finally, by performing an individualized false positive test on each light curve, thus testing the extent to which unique photometric variability of a given target may impact the moon classification for that star. We have seen that the CNN ensemble is capable of reaching up to 97\% accuracy in the training and validation stage. But do the simulated light curves in fact accurately reflect the systems we see in the real data?

There are three ways in which the (simulated) training and validation samples can differ from the real sample. First, the stellar activity could be different. Second, the planetary systems found in these light curves could be somehow fundamentally different in some key way. Finally, the processing of these light curves could be different. Let us handle these in turn:

\subsubsection{Stellar sample}
The ``donor'' light curve stars and the planet-hosting stars we vetted in this work differ in one obvious way: the latter are known to host planets, the former generally are not. Of course, this does not mean the donor stars do not host planets; rather, \textit{transiting} planets have not been discovered in this system. Because the chances of seeing a planet transit around a random star are geometrically and temporally quite low, and the estimated occurrence rate of planets is quite high, it is likely that many, indeed most, of the donor stars host non-transiting planets. There may of course also be small planets that are transiting but are so small that their transits are lost in the noise.

A detailed comparison of the stellar properties for light curves with and without planet discoveries is beyond the scope of this work. However, because we did not make any cuts on the KIC stars used as donor light curves (they were selected at random), the most obvious potential difference in the sample is likely the presence of some larger stars (early-type and giants) which are in general more challenging targets for exoplanet transit detection. On the other hand, the stars selected for monitoring by \textit{Kepler} were mostly hand-picked to be attractive targets for transit detection, and only a small fraction of the targets were O / B type stars ($< 200$) or giants ($\sim5000$) \citep{batalha:2010}. Thus we expect the donor light curve sample to largely reflect the planet-hosting sample. Once again, we expect that our pipeline effectively removed peculiar astrophysical signals through detrending and screened those targets with inadequate removal of systematics. And of course, our injected planet transits were large enough to be detectable, so in this respect they are not comparable to planets around giants, for which the primary difficulty is their extremely shallow transit depths.

\subsubsection{Ground truth architectures}
Until we begin to accumulate a number of exomoon detections in the literature, we cannot know for sure what the underlying population looks like. At present, we can only base our expectations on observations of the Solar System, and theoretical work, which, by and large, has been premised on the goal of explaining the satellites we find in our own Solar System. We should keep in mind, though, that systems unlike our Solar System may very well be out there and waiting to be found, so we should not restrict ourselves to systems we find \textit{plausible}. Any physically possible system should be in the sample, including large satellites. However, as previously mentioned, exceptionally large satellites were deemed to be unlikely to have evaded detection in the first place, and we therefore limited simulated moons radii to Earth sized (already on the large end of \textit{expected} moons, but considerably smaller than the exomoon candidate Kepler-1625b-i.). We further restricted our training sample to moons with single transit SNRs $\geq 5$. It is probably reasonable to assume that smaller moons, with lower SNRs, are hiding in the data. But if a moon is so small as to be lost in the noise, we have little hope of recovering it during model selection, nor can we expect to convince a skeptical community that a detection of such a low SNR object is real.

\subsubsection{Processing comparison}
The final possible difference between training and validation samples could be the processing pipeline. As we described above, we have made every effort to process them identically, using the same detrending algorithm and filtering approach for identifying short-duration variability and/or inadequate detrendings. The key difference between the two sets of light curves, then, could have to do with transit timings. 

In the simulated sample, we are able to place the planet transit precisely in the middle of the input window. For the real systems, we rely on established transit timings from the 
\citealt{holczer:2016} catalog where available, and assume linear ephemeris otherwise. To the extent that these could be inaccurate, placing some transits someplace other than the center of the segment, our samples could differ. Low SNR planets will generally have the worst transit timing errors, but these systems are already less desirable targets in the moon search. Future applications would be improved by data augmentation that more closely mimics the real sample by including planet transits that are off-center. It would also be worthwhile to explore the use of a training set that spaces moon transits uniformly in time, rather than using randomized orbit angles, to ensure all sky separations are equally well represented.

Inaccurate transit timings carry another effect, namely, the possibility of a badly masked transit. Badly masked transits can result in the introduction of artefacts that are potentially confusing to the CNN ensemble. An improperly masked transit may result in a light curve trend model that attempts to compensate for the transit itself, so that when the trend is divided out there is a bump in the light curve in the vicinity of the transit. Non-astrophysical transit depth changes may also occur, though this is less of a worry for the CNN because it analyzes individual transits and therefore does not know anything about depth changes from transit to transit. Any morphological manipulation that occurs from a bad detrending is of course undesirable, but once again this is generally restricted to low SNR planets and those systems with high frequency oscillations, which the CoFiAM detrending algorithm leaves alone.

\subsection{Fractional transit classifications}
A recurring issue in the identification of potential moon transits is that only in rare cases are 100\% of the analyzed transits for a given planet classified as having a moon transit present. Frequently, for systems here we might consider to be promising targets for follow-up, somewhere between 50\% and 75\% of the planet transits are classified as having a moon transit present also. What are we to make of this?

\subsubsection{System false positive probability}
Particularly when it comes to vetting long period planets, which we presume to be attractive targets in the moon search, we face a challenge; we simply lack a large number of transits with which to work. In terms of promising targets, how do we compare two planets, one with, say, 12 out of 20 transits showing a moon, versus one with 3 out of 5? They obviously have the same percentage of transits showing a moon feature. The shorter period planet has four times the number of moon predictions, and could therefore appear more promising; 3 could be a fluke, 12 seems less likely to be so. Can we quantify this \textit{system} false positive probability (FPP), beyond the planet-only injection test utilized previously?

We can attempt to compute this by appealing to the binomial theorem to compute a probability for this result arising from pure chance in these two scenarios, as a function of the probability $P(\mathrm{moon} \, | \, \mathrm{positive})$ that a single transit moon prediction is accurate. Here we use ``positive'' as shorthand for ``moon prediction''. We could let $P(\mathrm{moon} \, | \, \mathrm{positive})$ be the precision of the CNN ensemble, equal to the ratio of the number true positives to the number of all moon predictions (true positives and false positives). But this value does not account for the occurrence rate of moons. To incorporate this piece of information we could instead leverage Bayes' theorem to compute this as 

\begin{equation}
    P(\mathrm{moon} \, | \, \mathrm{positive}) = \frac{P(\mathrm{positive} \, | \, \mathrm{moon}) \, P(
    \mathrm{moon})} {P(\mathrm{positive})}.
\end{equation}

\noindent For the first term in the numerator we can insert the recall -- a tuneable parameter based on the agreement metric and desired precision, equal to the number of true positives divided by the number of moons. The denominator can be expanded as 

\begin{equation}
\begin{aligned}
    P(\mathrm{positive}) = \, & P(\mathrm{positive} \, | \, \mathrm{moon}) \, P(
    \mathrm{moon}) \, + \\
    & P(\mathrm{positive} \, | \, \mathrm{no  \, moon}) \, P(
    \mathrm{no \, moon})
    \end{aligned}
\end{equation}

\noindent where the probability $P(\mathrm{no \, moon}) = 1 - P(\mathrm{moon})$ as it is a binary choice. We can let $P(\mathrm{positive} \, | \, \mathrm{no  \, moon})$ be the global false positive rate from the training, but a more tailored approach would be to use the false positive rate calculated for each light curve based on the planet-only model injections. In any case, the difficulty here is that we cannot constrain the probability $P(\mathrm{moon})$ because the underlying occurrence rate is at present poorly constrained.

\begin{figure}
    \centering
    \includegraphics[width=\columnwidth]{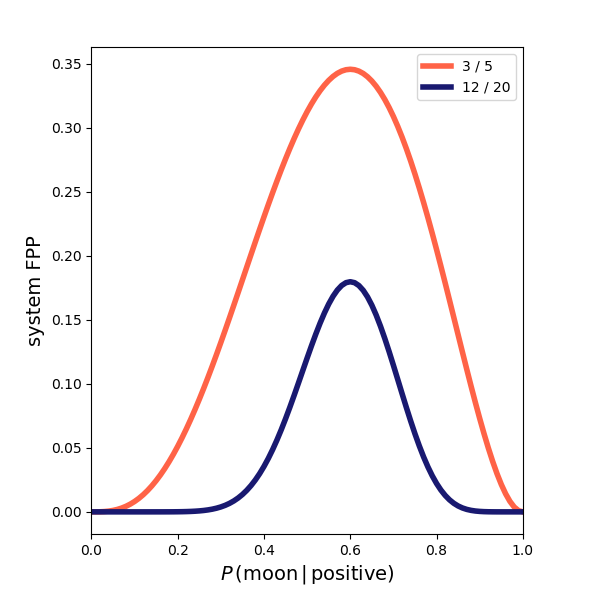}
    \caption{Chances of a false positive moon classification for two systems with a 60\% prediction rate, shown as a function of the chance moon prediction rate.}
    \label{fig:binomial}
\end{figure}

Returning to the problem of comparing two systems with a different number of transits but equal moon prediction percentages, we can plot the chances of each scenario as a function of the unknown $ P(\mathrm{moon} \, | \, \mathrm{positive})$. We display the full range of possibilities in Figure \ref{fig:binomial} for the two scenarios explored above (3/5 transits and 12/20 transits). Here the system FPP is the chance a system with no moons at all has gotten 3 out of 5 (or 12 out of 20) transits erroneously classified as containing a moon signal. Unsurprisingly, this probability peaks when $P(\mathrm{moon} \, | \, \mathrm{positive})$ matches the fraction of transits classified as having a moon. But importantly, the probability of a chance 60\% moon classification drops significantly the more transits we have vetted. At the wings of the distribution this probability drops very low.

In light of this, we can say that the longer period planets may be more attractive \textit{a priori} from the standpoint of physical considerations, but are less attractive targets in view of their follow-up opportunities and greater chance of being fluke moon classifications.

\subsubsection{Unknown system architectures}
A related problem concerns fractional moon classifications by virtue of the system's architecture. We would presumably have the highest confidence in a system that displays a moon transit in every epoch. Not only would the chances of a false positive be reduced by virtue of such consistency across all epochs, it would also accord better with our (perhaps naive) expectation that a moon should be more-or-less co-planar with the orbital plane of the planet. By contrast, a planet that does not register a moon transit during every epoch is worrisome. First, it suggests an inclined orbit, which makes it more challenging from a follow-up perspective. If the moon doesn't transit every epoch, it becomes difficult to argue for expensive telescope resources to be devoted, because a transit observation cannot be guaranteed -- though a putative moon's transit probability may be calculated \citep{martin:2019}. On the other hand, we do not have good reason to reject the moon hypothesis based on an inclined moon solution, and indeed, we see large inclinations in the Solar System with respect to the ecliptic, both because the moon itself is inclined (Triton), or the entire system is tipped over with respect to the orbital plane (Uranus).

Beyond this, we face a difficulty with the modeling. Because we cannot put a reasonable prior on the inclination of a moon (we do not know what they should be), and an inclined moon does not really constitute a more complex model, the model will not be penalized for placing a moon in such an orbit. Thus, a moon model can be flexible enough to ignore several, possibly even most, epochs that lack a moon signal, and it still accords with the data. The moon model may then be supported primarily by 1) the presence of dynamical effects, which do not have to be produced by a moon, and 2) by the occasional appearance of a moon-like dip.

\subsection{Future work}
Because an inclined moon is not a more complex model, and is therefore not penalized during model selection, what is needed instead is improved moon priors. At present we have good reason to be agnostic about features like the size and semimajor axis distribution, occurrence rate, the relative frequencies of prograde and retrograde orbits, the presence of resonances, and inclinations. Conditioning these solely on the Solar System, or on theoretical results, could bias us against unanticipated architectures. This is to say, to some extent we must begin to find exomoons so as to refine our priors for future exomoon searches -- a chicken and egg problem. Another useful line of research relates to the axial tilt of exoplanets. Predictions of an exoplanet's obliquity, or even better, direct observations of it, could be quite useful in tempering our expectations for the moon inclinations.

Despite the somewhat ambiguous results from the present study, CNNs, and other machine learning approaches, will remain attractive options in the search for exomoons in large data sets, and will certainly be worthwhile to apply to the K2 and TESS data sets. As these data pose their own unique challenges, we leave application of the tools developed here to other datasets for future work. 

In any case, the present results also demonstrate the continued need for rigorous model testing and close inspection of the results. Putative exomoon signals in the form of photometric dips are but one piece of the puzzle (the same can be said of using TTVs by themselves as evidence for ``candidate'' exomoons). We wish to develop tools that can help us identify quickly promising systems in a mountain of data, but we should expect that some, indeed perhaps many, of these systems may fail to meet our high standards for discovery, either due to insufficient data quality or because of some physical tension in the model results. At this stage, any peculiar or anomalous system feature is cause enough for concern and potentially rejection of the moon hypothesis. It is not that theses planets do not host moons; rather, the evidence is not strong enough (yet), and we must continue to look for more promising targets.

\section{Conclusions}
\label{sec:conclusions}

In this work we have carried out a systematic search for exomoons in the \kepler\ data, utilizing an ensemble of convolutional neural networks trained on a large sample of simulated data. We are able to achieve up to 97\% precision on the validation sample when the voting ensemble is in total agreement. We have applied this CNN ensemble to a total of 1880 KOIs with periods > 10 days (results available at \url{https://github.com/alexteachey/KOI_CNN_moon_results}), amounting to $\sim$ 57,000 transit events. Of these, only about 1.5\% were classified as having transit events, and a mere 0.3\% were classified as containing moon transits after our agreement metric threshold was applied. While we caution the reader against strong conclusions about the occurrence rate of exomoons based on this result, it does suggest that detectable moon transits will remain challenging to identify in the near future. 

Broadly speaking, our efforts in ensuring the training sample closely resembles the sample of real KOIs were aimed primarily at suppressing an anticipated large  number of false positives. However, it is clear in the final analysis that an overabundance of moon transit predictions is not the problem at all. We have come up with very few such predictions, and while such a result in unanticipated, we are obliged to trust the accuracy of the CNN ensemble based on its performance with the simulated dataset. Unfortunately it is just not yet possible to validate the performance of the CNN on real, known exomoon systems.

This work highlights some of the ongoing challenges with identifying exomoons, and in particular, the difficulty of applying a machine learning framework to a problem with so many lingering unknowns. While the experimenter may be confident in the ability of machines to correctly classify simulated light curves through training and validation, application to real data sets poses some additional challenges. The unique features of each light curve, and underlying ignorance of everything happening in the system, can lead to predictions that require the individualized scrutiny which the machine learning approach is intended to avoid. Old-fashioned vetting, including checks on the physicality and plausibility of moon solutions, remains important.

Nevertheless, CNNs are and should remain a useful framework for approaching the problem of identifying candidate exomoon signals in large data sets, and we intend to extend this work to other large surveys in the near future, continuing to refine the methodology in the process. Additional machine learning approaches may also be leveraged, but we will still require diligent and sober analysis of these targets.

\section*{Acknowledgements}
We thank the anonymous reviewers for their constructive feedback, which significantly improved this work. A sizeable fraction of this work was performed while AT was a PhD candidate at Columbia University, where he was supported through the NSF Graduate Research Fellowship
(DGE-1644869). DK acknowledges support from NASA Exoplanet Research Program grant number 80NSSC21K0960. The Cool Worlds Lab is supported by donors including Tom Widdowson, Mark Sloan, Laura Sanborn, Douglas Daughaday, Andrew Jones, Elena West, Tristan Zajonc, Chuck Wolfred, Lasse Skov, Alex de Vaal, Jason Patrick-Saunders, Methven Forbes, Stephen Lee, Zachary Danielson, Vasilen Alexandrov, Chad Souter, Marcus Gillette, Tina Jeffcoat, Jason Rockett, Scott Hannum, Tom Donkin and Mark Elliott.

This paper includes data collected by the Kepler mission. Funding for the Kepler mission is provided by the NASA Science Mission directorate. This research has made use of the NASA Exoplanet Archive, which is operated by the California Institute of Technology, under contract with the National Aeronautics and Space Administration under the Exoplanet Exploration Program. This research has made use of NASA's Astrophysics Data System. We also gratefully acknowledge the developers of the following software packages which made this work possible: \texttt{Astropy} \citep{astropy}, \texttt{NumPy} \citep{numpy1, numpy2}, \texttt{Matplotlib} \citep{matplotlib}, \texttt{Keras} \citep{keras}, \texttt{Tensorflow} \citep{tensorflow}, and \texttt{Corner} \citep{corner}.  

\section*{Data Availability}
The results of our CNN ensemble KOI vetting are available at \url{https://github.com/alexteachey/KOI_CNN_moon_results}. Much of the this analysis was carried out using tools within the \texttt{MoonPy} package developed by AT, available at \url{https://github.com/alexteachey/MoonPy}. These tools are frequently updated and may no longer reflect their state at the time of this analysis. Additional inquiries regarding the data or analysis may be directed to the corresponding author.

%%%%%%%%%%%%%%%%%%%%%%%%%%%%%%%%%%%%%%%%%%%%%%%%%%



%\bibliographystyle{mnras}


%%%%%%%%%%%%%%%%%%%%%%%%%%%%%%%%%%%%%%%%%%%%%%%%%%

%%%%%%%%%%%%%%%%% APPENDICES %%%%%%%%%%%%%%%%%%%%%

%\appendix



%%%%%%%%%%%%%%%%%%%%%%%%%%%%%%%%%%%%%%%%%%%%%%%%%%

% Don't change these lines
\bsp	% typesetting comment
\label{lastpage}
\end{document}